\documentclass[twocolumn,showpacs,amsmath,amssymb]{revtex4-1}

\usepackage{graphicx}
\usepackage{dcolumn}
\usepackage{bm}
\usepackage{xcolor}
\def\be{\begin{equation}}
\def\ee{\end{equation}}
\def\ba{\begin{array}}
\def\ea{\end{array}}
\def\bea{\begin{eqnarray}}
\def\eea{\end{eqnarray}}
\begin{document}

\title{Systematic study of complete fusion suppression for $^{6,7}$Li nuclei using artificial neural network}
\author{ D. Chattopadhyay$^{1}$}
\email{dipayanchattopadhyay90@gmail.com}
\affiliation{$^1$Department of Physics, The ICFAI University Tripura, Agartala 799210, India}
\date{\today}

\begin{abstract}
In recent decades, there has been a significant increase in the measurement of complete fusion cross-sections for various reactions, with particular emphasis on the reactions involving weakly bound projectile. It has been well established that the complete fusion cross-section involving weakly bound nuclei is suppressed at above barrier energies due to the breakup effect. Accurate determination of suppression factor is essential to understand the effect of breakup on complete fusion suppression. In this study, Feedforward Artificial Neural Network (ANN) methods based on Multilayer Perceptron is utilized to estimate the complete fusion suppression factor for reactions involving $^{6,7}$Li projectile from the comparison of ANN predicted reduced fusion functions ($F(x)$) with the Universal Fusion Function($F_{0}(x)$). Average suppression factor has been estimated as 0.68 and 0.74 for $^{6}$Li and $^{7}$Li induced reactions respectively. Normalized Mean Squared Error(NMSE) has been calculated as 1.85$\%$ for training and 1.92$\%$ for testing data of ANN for $^{6}$Li case, while for $^{7}$Li cases, the same is for 3.73$\%$ and 6.48$\%$ respectively. Present results have been compared with three other alternative methods Support Vector Regression, Random Forest Regression and Gaussian Process Regression.  Relevant performance matrices has been estimated. It has been observed that, ANN algorithm is giving the better accuracy as compared with other. This results indicate that ANN might be an alternative tool for estimation of complete fusion suppression factor for weakly bound projectile.        
\end{abstract}

\maketitle

\section*{Introduction}
Nuclear reactions occurring near the Coulomb barrier serve as an extensive repository revealing diverse facets of fundamental quantum mechanics. This phenomenon becomes even more intricate if the projectile is weakly bound in nature. In nuclear systems with weak binding, correlations among nucleons and pairing give rise to phenomena such as the emergence of strong clustering and exotic shapes. Lithium isotopes such as $^{6,7}$Li offer a distinctive illustration of nuclear clustering, characterized by a well-known $\alpha$ + $x$ cluster structure. Indeed, $^{6}$Li exhibits a well-known $\alpha$ + $d$ cluster structure, featuring a breakup threshold of approximately 1.47 MeV. Similarly, $^{7}$Li is characterized by a well-established $\alpha$ + $t$ cluster configuration, with a breakup threshold of around 2.47 MeV. The investigation of nuclear reactions involving weakly bound projectiles $^{6,7}$Li has garnered significant interest, driven by the observation of numerous novel features distinct from those associated with strongly bound projectiles. Suppression in complete fusion (CF) cross sections~\cite{tripathi02,liu05,shrivastava09,dasgupta04,rath09}, breakup threshold anomaly in the optical potentials obtained from elastic scattering~\cite{santra11}, and high yield in $\alpha$ particle production~\cite{chattopadhyay16,chattopadhyay18} are some of the intriguing features in this context. The existence of projectile breakup channels, along with other nonelastic channels, and their interaction with the elastic channel, are the key factors contributing to the aforementioned distinctions. Numerous measurements of complete fusion (CF) cross sections have been conducted~\cite{tripathi02,liu05,shrivastava09,dasgupta04,rath09}. 

Various methods have been employed to explore the influence of breakup on fusion reactions around the Coulomb barrier~\cite{hagino00,diaz02,gomes11}. One widely used approach is to compare experimental data either with predictions from coupled channel (CC) calculations that do not consider breakup channels~\cite{kumawat12,rath12,pradhan11,palshetkar14,rath13} or with the predictions of a single barrier penetration model (SBPM)~\cite{tripathi02,tripathi05}. Observations indicate that complete fusion (CF) cross sections are suppressed at energies above the Coulomb barrier. A novel approach proposed in Ref.~\cite{canto09} involves assessing the dynamic impact of projectile breakup on fusion by comparing a dimensionless reaction fusion function derived from experimental data with a theoretical system-independent universal fusion function (UFF). In this method, experimental fusion cross sections are normalized to theoretical fusion cross sections obtained from coupled-channels calculations, incorporating only the pertinent bound inelastic and transfer channels. Any disparity between the two fusion functions is attributed to the effect of projectile breakup. However, it is worth noting that the degree of the suppression factor obtained through the aforementioned approaches is observed to be contingent on the specific modeling choices employed~\cite{kundu16}.   

Artificial neural networks (ANN) serve as a powerful mathematical tool for estimating values across various fields in science and technology, including nuclear physics studies. This method is particularly effective in providing accurate results for highly nonlinear relationships between dependent and independent data points. Recently, ANN has found application in numerous areas of nuclear physics, such as constructing consistent physical formulas for detector counts in neutron exit channel selection~\cite{akkoyun13}, determining one and two proton separation energies~\cite{athanassopoulos04}, developing nuclear mass systematic~\cite{athanassopoulos004}, determining ground state energies of nuclei~\cite{bayram14}, identifying impact parameters in heavy-ion collisions~\cite{david95,bass96,haddad97}, determining beta-decay energies~\cite{akkoyun14}, and estimating nuclear rms charge radius~\cite{akkoyun013}.   

Understanding the dynamical effect of fusion suppression through theoretical formulation requires many physical factors like location of breakup, coupling mechanism between breakup channel with the fusion channel. In machine learning approaches for predicting cross-sections, the goal is not to understand or define the physical mechanisms behind the fusion reaction but to achieve the most accurate cross-section predictions. This means that cross-section information can be obtained without relying on any physical formulations. If only cross-section information is required, this is an advantage of using machine learning. By training on existing experimental data, the machine can generate new data independently of physical formulations. Since the most reliable cross-section information comes from experiments, the potent tool of Artificial Neural Networks (ANN)~\cite{haykin99} is utilized in present investigation to acquire complete fusion cross-section data involving of $^{6,7}$Li. The experimental datas were extracted from existing literature~\cite{tripathi02,liu05,shrivastava09, dasgupta04,rath09,tripathi05,kumawat12,rath12,pradhan11,palshetkar14,rath13,mukherjee06,shrivastava13}.

In the present work, the data sets consisting of dimensionless variable $x$[=$\dfrac{E_{c.m}-V_{B}}{\hbar\omega}$] as an input and Fusion Function F(x) as output, where E$_{cm}$ is the energy of the incident projectile in c.m. frame, V$_{B}$ is the fusion barrier and $\hbar$$\omega$ is the width of the barrier. The fusion barrier is nothing but the combination of coulomb barrier, nuclear barrier and centrifugal barrier between the projectile and the target. It is clear that, mass of the target nucleus is explicitly there inside the expression of fusion barrier. Hence the effect is target mass is already been included here. Through training the ANN with known Fusion Function F(x) with $x$, the relationship between F(x) with $x$ is estimated. Any deviation of ANN predicted Fusion Function F(x) from the UFF should be attributed to clustering effects and the intensity of the effect could be estimated by the strength of the deviation. For that, the average suppression factor for both $^{6}$Li and $^{7}$Li induced reactions, has been calculated using the ratio of ANN model predicted F(x) with UFF(x). The present result is also compared with the other alternative interpolating methods Support Vector Regression, Random Forest Regression and Gaussian Process Regression. 
The details of Artificial Neural Network method will be discussed in next section followed by Calculations of fusion suppression factor, Results and Discussion and Conclusions.

\section*{Artificial Neural Network}
An artificial neural network (ANN) is a computational model inspired by the structure and function of biological neural networks in the human brain. It is a key component of machine learning and artificial intelligence systems, and it's designed to mimic the way the human brain processes information. Neurons are the basic building blocks of artificial neural networks. Each neuron receives one or more inputs, performs a computation, and produces an output. Neurons are organized into layers: an input layer, one or more hidden layers, and an output layer. If information flows in one direction: from the input layer to the output layer without cycles or loops, then the ANN is named as feed-forward neural network. The network consists of multiple layers of nodes (neurons). It has at least one hidden layer between the input and output layers. Neurons are connected by edges, each associated with a weight. The weight represents the strength of the connection between neurons. A bias term is also often used to shift the output of a neuron. During training, the weights and biases are adjusted to optimize the network's performance. The activation function determines the output of a neuron given its input. It introduces non-linearity into the network, allowing it to learn complex patterns. The network has a single output node, suitable for regression tasks, as the goal is to predict a continuous variable. During the feed-forward operation, the input data is propagated through the network, layer by layer. The output of each neuron is computed based on the weighted sum of its inputs and the application of the activation function. The training of a feed-forward neural network involves adjusting the weights and biases based on the difference between the predicted output and the actual target. This is typically done using a supervised learning approach and an optimization algorithm, such as gradient descent. The gradients of the loss with respect to the weights and biases are calculated. This involves computing how much the loss would change if each weight and bias were adjusted. The weights and biases are adjusted in the direction that minimizes the loss, typically using an optimization algorithm like stochastic gradient based descent, Adam(Adaptive Moment Estimation) etc. In the present analysis, grid-search will explore both `Adam' and `Stochastic Gradient Desecent' and the best-performing solver will be determined during the hyperparameter tuning process. The above processes are repeated for a specified number of training epochs (maxiter) to iteratively improve the model.

The mathematical formulation of a simple feedforward neural network (Multi-Layer Perceptron) used for regression is discussed in the following:
Layer by layer, the breakdown will be initiated, followed by the introduction of the training process involving gradient descent.
\subsection*{Forward Pass:}
\begin{figure}[h]
\includegraphics[scale=0.40]{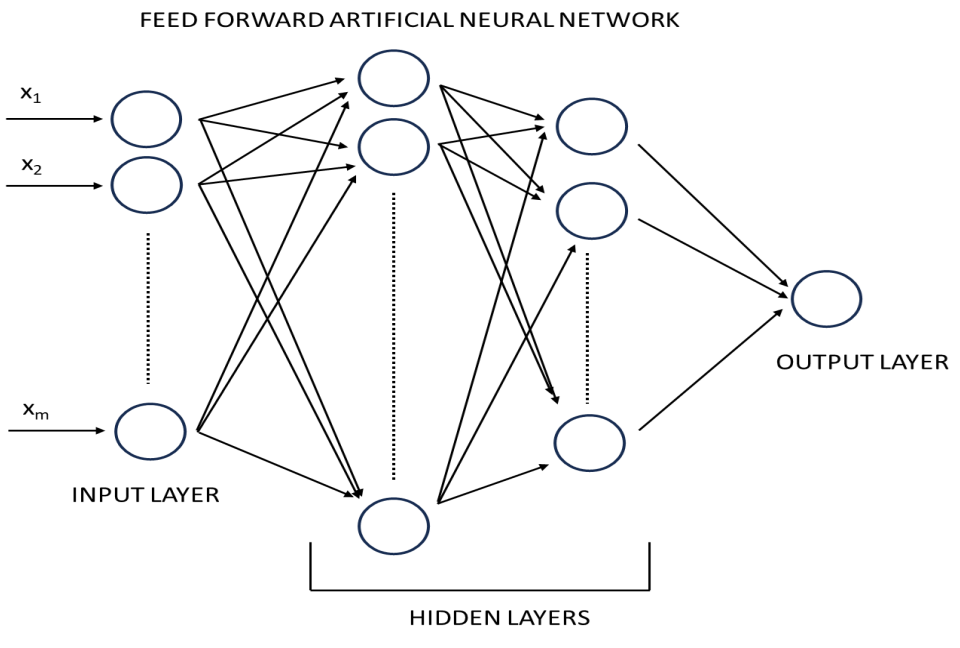}
\caption{\label{fig:ann} Typical architecture of Feed Forward Neural Network }
\end{figure}
\subsection{Input Layer:}
The input to the neural network is a vector X representing the features of a single data point. If there are m features, X is a column vector of size m $\times$ 1, often represented as 
\begin{align}
    X &= \begin{bmatrix}
           x_{1} \\
           x_{2} \\
           \vdots \\
           x_{m}
         \end{bmatrix}
  \end{align}   
Here, each x$_{i}$ represents a feature of the input data point. The information (x$_{1}$,x$_{2}$,...,x$_{m}$) passes to the nodes in the hidden layer. 

\subsection{Hidden Layer:}
For a single hidden layer with n neurons, the output Z$_{hidden}$ is computed as:
\begin{equation}
Z_{hidden}=(W_{hidden}\cdot X+b_{hidden})
\end{equation} 
where,

\begin{figure*}
\centering\includegraphics[scale=0.80]{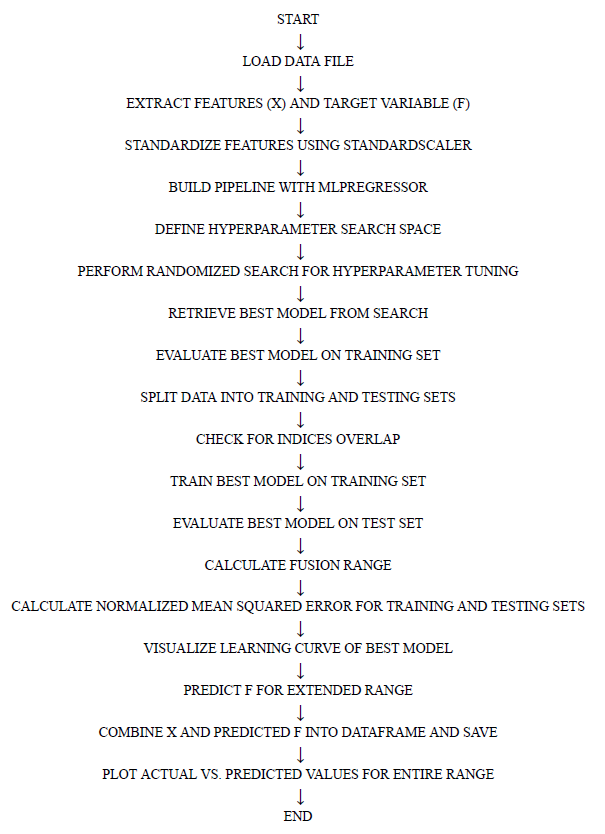}
\caption{\label{fig:flowchart} Flowchart of the ANN model}
\end{figure*}

W$_{hidden}$ is the weight matrix of size n $\times$ m.

b$_{hidden}$ is the bias vector of size n $\times$ 1.

After computing Z$_{hidden}$, it is passed through an activation function $\sigma$:
\begin{equation}
A=\sigma(Z_{hidden})
\end{equation}
where A is the output(activation) of the hidden layer. If the weighted sum of the inputs to a neuron is above the the threshold or activation layer, then the neuron fires. In this analysis, the activation function for the hidden layers is being explored with two options: `ReLU' (Rectified Linear Unit) and `tanh' (Hyperbolic Tangent). The model will be trained and evaluated with both activation functions to determine which performs better for the given data.

So, the overall computation can be written as:
\begin{equation}
A=\sigma(W_{hidden}\cdot X+b_{hidden})
\end{equation}

This represents the forward propagation through the first hidden layer of a neural network. The activation function $\sigma$ introduces non-linearity to the model, allowing it to learn complex patterns in the data.

If there are two hidden layers in the neural network, the computation of the output at the first hidden layer (Z$_{hidden1}$) and the subsequent hidden layer (Z$_{hidden2}$) can be expressed as follows:

For the first hidden layer:
\begin{equation}
Z_{hidden1}=\sigma(W_{hidden1}\cdot X+b_{hidden1})
\end{equation} 
Here:

    W$_{hidden1}$ is the weight matrix of the first hidden layer with dimensions n$_{1}$ $\times$ m,
    
    b$_{hidden1}$ is the bias vector of the first hidden layer with dimensions n$_{1}$ $\times$ 1,

    $X$ is input vector with dimensions $m$ $\times$ 1, 
    
    $\sigma$ represents the activation function already mentioned above.

For the second hidden layer:
\begin{equation}
Z_{hidden2}=\sigma(W_{hidden2}\cdot Z_{hidden1}+b_{hidden2})
\end{equation} 

Here:

    W$_{hidden2}$ is the weight matrix of the second hidden layer with dimensions n$_{2}$ $\times$ n$_{1}$,
    
    b$_{hidden2}$ is the bias vector of the second hidden layer with dimensions n$_{2}$ $\times$ 1.

    Z$_{hidden1}$ is output of the first hidden layer with dimensions n$_{1}$ $\times$ 1.  

This process extends the neural network's capacity to learn hierarchical features and complex representations from the input data.

\subsection{Output Layer:}
For regression, there is typically no activation function in the output layer, and the output Y is given by:
\begin{equation}
Y=W_{output} \cdot Z_{hidden2}+b _{output}
\end{equation} 

where, W$_{output}$ is a weight matrix of size 1 $\times$ n$_{2}$

Z$_{hidden}$ is the output from the second hidden layer,

b$_{output}$ is the bias term for the output layer.

\subsection*{Training with Stochastic Gradient Descent:}

\subsection{Loss Function:}

The Mean Squared Error (MSE) is commonly used for regression:
\begin{equation}
MSE=\dfrac{\sum_{i=1}^{N} (\hat{Y_{i}}-Y_{i})^2}{N}
\end{equation} 
where,
N is the data points.
$\hat{Y_{i}}$ is the predicted output for data point $i$.
Y$_{i}$ is the actual output for data point i.
Normalized Mean Squared Error(NMSE) is defined as:
\begin{equation}
NMSE=\dfrac{MSE}{F_{max}-F_{min}}
\end{equation} 
where, F$_{max}$ and F$_{min}$ is the maximum value of $F(x)$ and minimum value of $F(x)$.
 In addition to that, the regularization term $R$ is also added to the loss function during training to prevent over-fitting by penalizing large weights in the model. For L2 regularization, the regularization term is proportional to the square of the weights:
\begin{equation}
R_{L2}=\dfrac{\alpha}{2}\sum_{i=1}^{n} W_{i}^{2}
\end{equation} 
Here, $\alpha$ is the regularization parameter, controls the strength of the regularization effect. Higher values of $\alpha$ result in stronger regularization.
\subsection{Back Propagation:}

Backpropagation (short for "backward propagation of errors") is a supervised learning algorithm commonly used for training artificial neural networks. It is a gradient-based optimization algorithm that minimizes the error between the predicted output and the actual target output. The key steps involved in the ANN process are:

(1) Compute the gradient of the regularization term with respect to the output layer weights: $\dfrac{\partial R}{\partial W_{output}}$.

(2) Compute the gradient of the mean squared error with respect to the output layer weights: $\dfrac{\partial MSE}{\partial W_{output}}$.

(3) Compute the gradient of the mean squared error with respect to the output layer biases: $\dfrac{\partial MSE}{\partial b_{output}}$.

(4) Propagate the gradients through the network to compute the gradient with respect to the second hidden layer weights and biases: $\dfrac{\partial R}{\partial W_{hidden2}}$,$\dfrac{\partial MSE}{\partial W_{hidden2}}$, $\dfrac{\partial MSE}{\partial b_{hidden2}}$ .

(5) Propagate the gradients through the network to compute the gradient with respect to the first hidden layer weights and biases: $\dfrac{\partial R}{\partial W_{hidden1}}$,$\dfrac{\partial MSE}{\partial W_{hidden1}}$, $\dfrac{\partial MSE}{\partial b_{hidden1}}$ .

The process involves backward propagation of the gradients through each layer of the network. The chain rule is applied to compute the gradients at each layer based on the gradients computed in subsequent layers.

\subsection{Update Weights and Biases:}

Update the weights and biases using the gradients and a learning rate $\eta$:
\begin{equation}
W_{output} \leftarrow W_{output}-\eta(\alpha\dfrac{\partial R}{\partial W_{output}}+\dfrac{\partial MSE}{\partial W_{output}})
\end{equation} 
\begin{equation}
b_{output} \leftarrow b_{output}-\eta \dfrac{\partial MSE}{\partial b_{output}}
\end{equation} 
\begin{equation}
W_{hidden2} \leftarrow W_{hidden2}-\eta(\alpha\dfrac{\partial R}{\partial W_{hidden2}}+\dfrac{\partial MSE}{\partial W_{hidden2}})
\end{equation} 
\begin{equation}
b_{hidden2} \leftarrow b_{hidden2}-\eta \dfrac{\partial MSE}{\partial b_{hidden2}}
\end{equation} 
\begin{equation}
W_{hidden1} \leftarrow W_{hidden1}-\eta(\alpha\dfrac{\partial R}{\partial W_{hidden1}}+\dfrac{\partial MSE}{\partial W_{hidden1}})
\end{equation} 
\begin{equation}
b_{hidden1} \leftarrow b_{hidden1}-\eta \dfrac{\partial MSE}{\partial b_{hidden1}}
\end{equation} 
where, $R$ represents the regularization term added to the loss function during training. 

This process is repeated for maximum iterations until convergence.
The training with Stochastic Optimization using Adam(Adaptive Moment Estimation) has been described in ~\cite{kingma14}. 
In the present calculations, the data sets consisting of dimensionless variable $x$ as an input and Fusion Function F(x) as output. For $^{6}$Li case, There are 114 data points and hence the size of the input x is 114 X 1. Whereas, for $^{7}$Li case, the data points are 80 and hence the size of the input here is 80 X 1. In the network structure, there are 1 input node in the input layer, 64 nodes in the hidden layer 1, 32 nodes in the hidden layer 2 and 1 node in the output layer. The number of hidden layer and hidden units are so chosen to get the minimum loss as much as possible. Hence, The total number of adjustable weights are 2144 and the biases are 97. Therefore, the total number of network parameters are 2241. The maximum iteration is used as 5000. In the Adam algorithm, the initial learning rate is 0.0001 and decay constants are 0.8 and 0.98. 

In the present case, Parameter grid search is employed in the provided code to fine-tune the hyperparameters of the MLPRegressor model. The grid search space encompasses various hyperparameters such as the hidden layer sizes, activation function, alpha (regularization parameter), learning rate, solver, and others. Each combination of hyperparameters is systematically evaluated using cross-validation with five folds. RandomizedSearchCV from scikit-learn is utilized for this purpose, with 20 iterations to randomly sample hyperparameter configurations from the specified parameter grid. This approach allows for an exhaustive search over the hyperparameter space while mitigating the computational cost associated with evaluating all possible combinations. By combining regularization techniques with cross-validation during grid search, the code aims to find the optimal hyperparameters that not only minimize the training error but also promote good generalization performance on unseen data. This comprehensive approach helps in addressing overfitting by ensuring that the selected model is robust and has good predictive power across different data samples.

\begin{table} [h]
\caption{Data table along with experimental suppression factor and ANN-predicted suppression factor}
\label{tab1}
\vspace*{0.1cm}
\begin{tabular}{ccccc}\hline \hline
&&&&\\
Proj. & Target & F$_{Bu.}$ & F$_{Bu.}$ \\
&     &  [expt.] & [ANN] \\
&&&& \\ \hline
&&&& \\
$^{6}$Li&$^{90}$Zr~\cite{kumawat12}&0.66$\pm$0.08&0.68\\
&&&& \\
$^{6}$Li&$^{144}$Sm~\cite{rath09}&0.68&0.68\\
&&&& \\
$^{6}$Li&$^{152}$Sm~\cite{rath12}&0.72$\pm$0.04&0.68\\
&&&& \\
$^{6}$Li&$^{159}$Tb~\cite{pradhan11}&0.66$\pm$0.05&0.68\\
&&&& \\
$^{6}$Li&$^{197}$Au~\cite{palshetkar14}&0.65$\pm$0.23&0.68\\
&&&& \\
$^{6}$Li&$^{198}$Pt~\cite{shrivastava09}&      &0.68\\
&&&& \\
$^{6}$Li&$^{208}$Pb~\cite{liu05}&      &0.68\\
&&&& \\
$^{6}$Li&$^{209}$Bi~\cite{dasgupta04}&0.66$\pm$0.05&0.68\\
&&&& \\
$^{7}$Li&$^{144}$Sm~\cite{rath13}&0.75$\pm$0.04&0.74\\
&&&& \\
$^{7}$Li&$^{152}$Sm~\cite{rath13}&0.75$\pm$0.04&0.74\\
&&&& \\
$^{7}$Li&$^{159}$Tb~\cite{mukherjee06}&0.74&0.74\\
&&&& \\
$^{7}$Li&$^{165}$Ho~\cite{tripathi02,tripathi05}&0.70&0.74\\
&&&& \\
$^{7}$Li&$^{197}$Au~\cite{palshetkar14}&0.85$\pm$0.04&0.74\\
&&&& \\
$^{7}$Li&$^{198}$Pt~\cite{shrivastava13}&       &0.74\\
&&&& \\
$^{7}$Li&$^{209}$Bi~\cite{dasgupta04}&0.74$\pm$0.03&0.74\\
&&&& \\ \hline
\end{tabular}
\end{table}

\begin{table} [h]
\caption{Comparison of performance matrix of different machine learning regression models for $^{6}$Li induced reactions}
\label{tab2}
\vspace*{0.0cm}
\begin{tabular}{ccccccc}\hline \hline
&&&&\\
Model & F$_{Bu.}$ & R$^{2}$ & R$^{2}$ & NMSE & NMSE \\
&  [Model] & [Training] & [Testing] & [Training] & [Testing] \\
&&&& \\ \hline
&&&& \\
ANN&0.680&0.98&0.98&1.85$\%$ &1.92$\%$\\
&&&& \\
GPR&0.674&0.95&0.93&2.3$\%$ &3.70$\%$\\
&&&& \\
SVR&0.701&0.975&0.96&1.93$\%$ &2.28$\%$\\
&&&& \\
RFR&0.677&0.986&0.936&1.18$\%$ &3.54$\%$\\
&&&& \\ \hline
\end{tabular}
\end{table}

\begin{table} [h]
\caption{Comparison of performance matrix of different machine learning regression models for $^{7}$Li induced reactions}
\label{tab3}
\vspace*{0.0cm}
\begin{tabular}{ccccccc}\hline \hline
&&&&\\
Model & F$_{Bu.}$ & R$^{2}$ & R$^{2}$ & NMSE & NMSE \\
&  [Model] & [Training] & [Testing] & [Training] & [Testing] \\
&&&& \\ \hline
&&&& \\
ANN&0.740&0.970&0.952&3.73$\%$ &6.48$\%$\\
&&&& \\
GPR&0.752&0.968&0.962&3.77$\%$ &5.18$\%$\\
&&&& \\
SVR&0.720&0.960&0.950&3.91$\%$ &6.61$\%$\\
&&&& \\
RFR&0.747&0.980&0.923&2.66$\%$ &7.93$\%$\\
&&&& \\ \hline
\end{tabular}
\end{table}
\begin{figure}[h]
\includegraphics[scale=0.40]{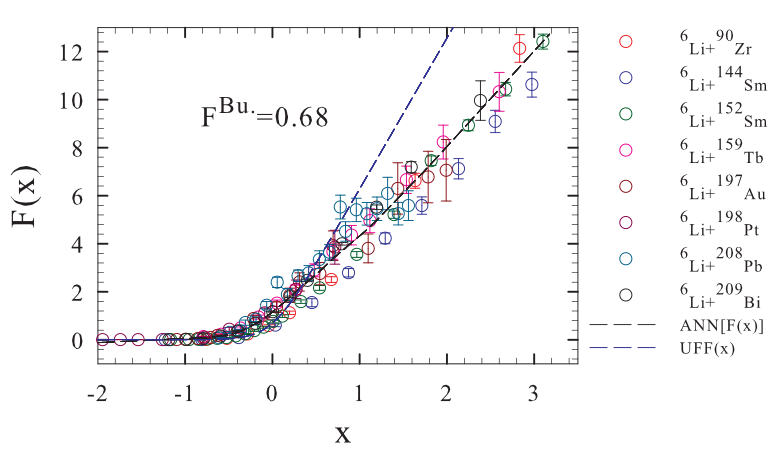}
\caption{\label{fig:6Li} Variation of complete fusion function F(x) with x for $^{6}$Li projectile on different target nuclei. }
\end{figure}
\begin{figure}[h]
\includegraphics[scale=0.40]{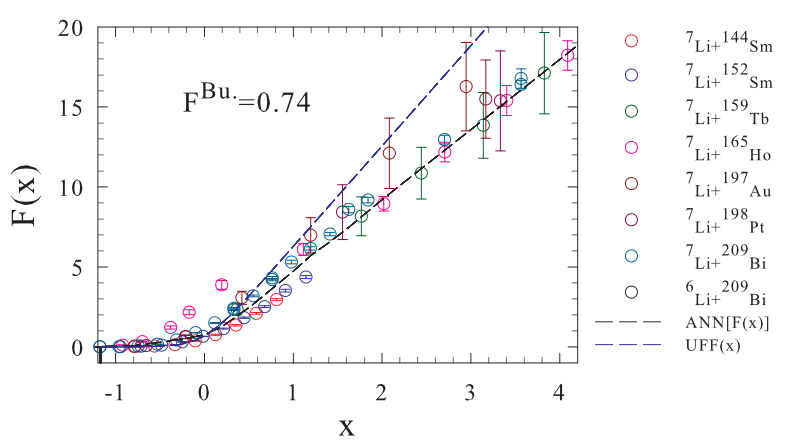}
\caption{\label{fig:7Li} Variation of complete fusion function $F(x)$ with $x$ for $^{7}$Li projectile on different target nuclei. }
\end{figure}

\begin{figure}[h]
\includegraphics[scale=0.40]{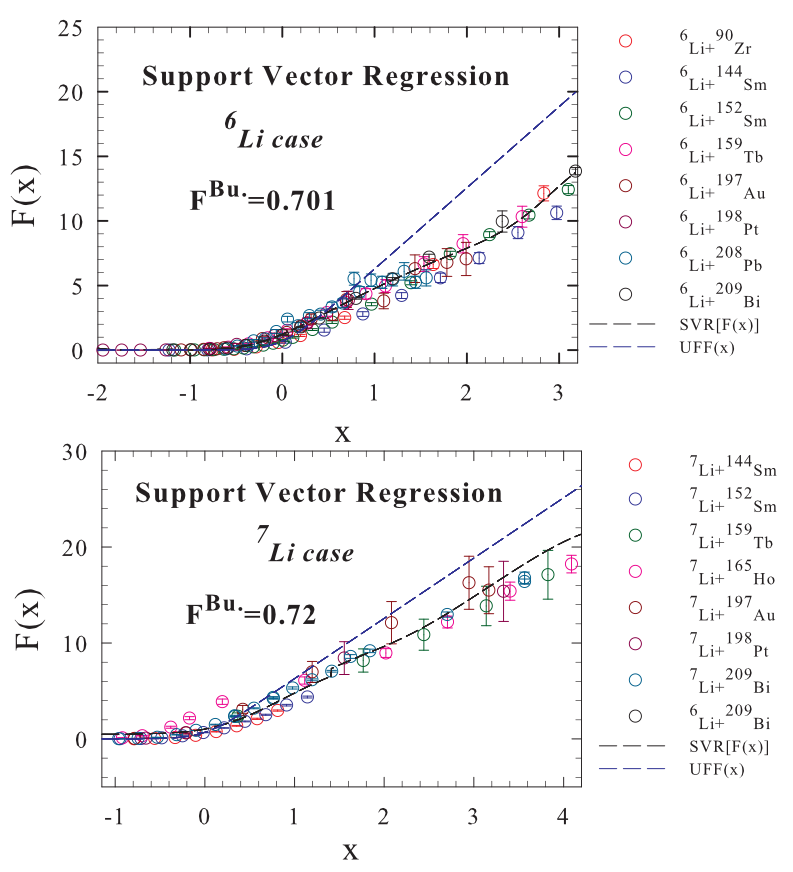}
\caption{\label{fig:svr} Variation of complete fusion function $F(x)$ with $x$ for $^{6}$Li[top panel] and $^{7}$Li[bottom panel] projectile on different target nuclei for Support Vector Regression.} 
\end{figure}

\begin{figure}[h]
\includegraphics[scale=0.40]{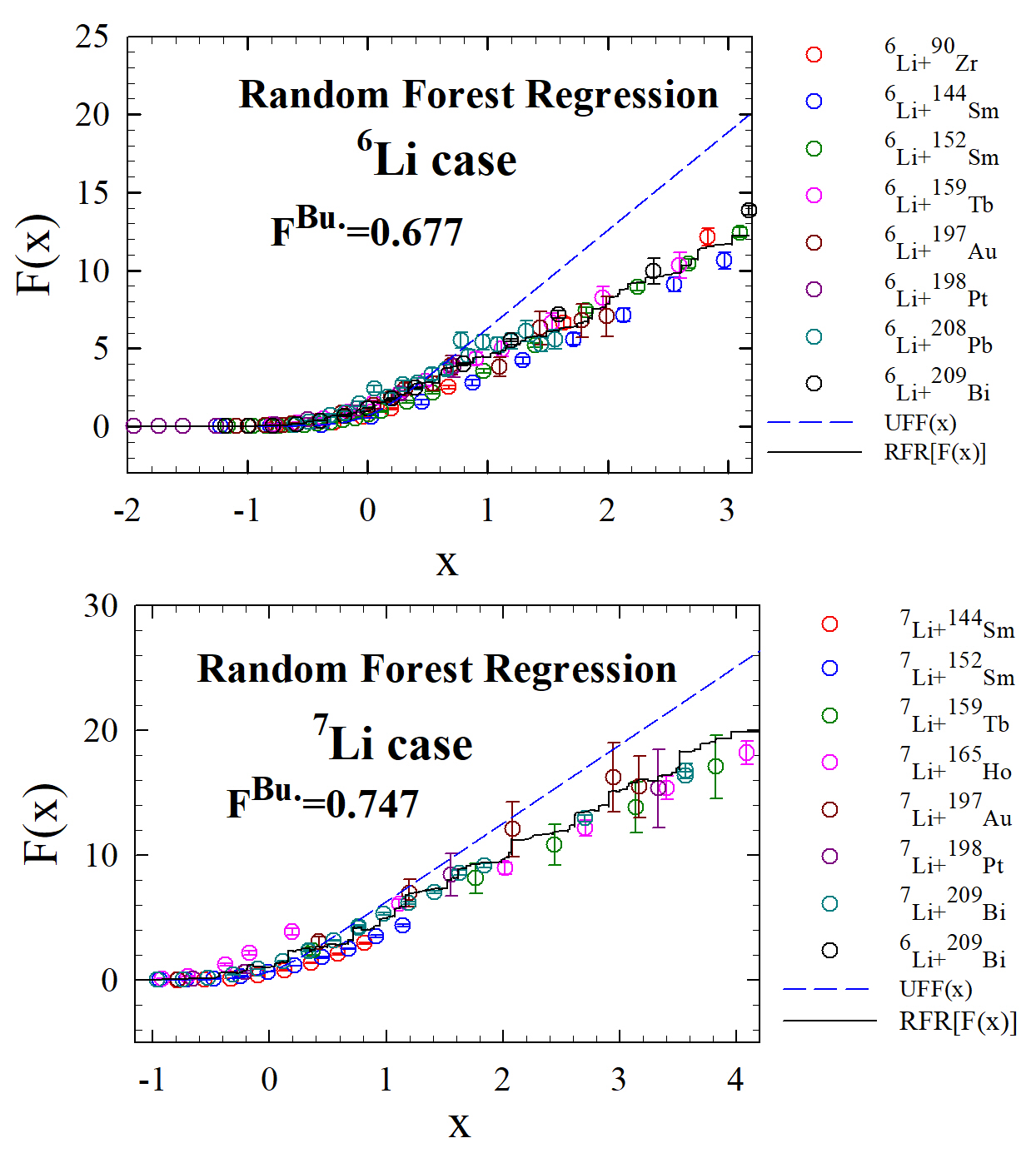}
\caption{\label{fig:rfr} Variation of complete fusion function $F(x)$ with $x$ for $^{6}$Li[top panel] and $^{7}$Li[bottom panel] projectile on different target nuclei for Random Forest Regression.} 
\end{figure}

\begin{figure}[h]
\includegraphics[scale=0.40]{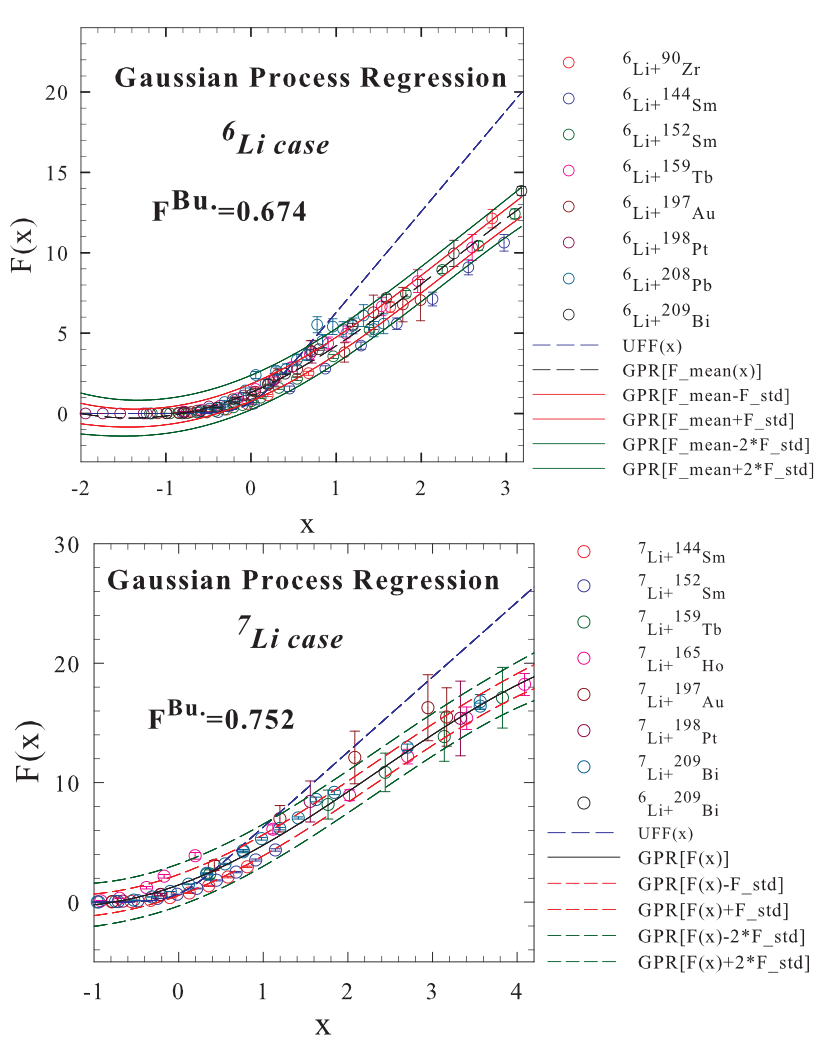}
\caption{\label{fig:gpr} Variation of complete fusion function $F(x)$ with $x$ for $^{6}$Li[top panel] and $^{7}$Li[bottom panel] projectile on different target nuclei for Gaussian Process Regression.} 
\end{figure}

\section*{Calculations of Fusion Suppression Factor}
To investigate the systematic patterns of breakup effects on CF cross sections, it is imperative to entirely eliminate the geometric factors and static influences arising from the potential between the two nuclei. According to the method proposed by Canto {\it et al.}~\cite{canto09}, the fusion cross section and collision energy are transformed into a dimensionless fusion function, denoted as $F(x)$, and a dimensionless variable, $x$.
\begin{equation}
F(x)=\dfrac{2E_{c.m.}}{R_{B}^{2}\hbar\omega}\sigma_F,    x=\dfrac{E_{c.m}-V_{B}}{\hbar\omega}
\end{equation} 
where, E$_{c.m}$ is the energy of incident projectile in the c.m. frame, $\sigma_{F}$ is the fusion cross-section and R$_{B}$,V$_{B}$ and $\hbar\omega$ implies the radius, fusion barrier and curvature of the barrier which is approximated as parabolic in nature. The barrier parameters R$_{B}$,V$_{B}$ and $\hbar\omega$ are calculated from Single Barrier Penetration model using NRV~\cite{nrv}.

According to the prescription, of Wong, the fusion cross-section for parabolic barrier is given by
\begin{equation}
\sigma^{Wong}_{F}=\dfrac{\hbar\omega R_{B}^{2}}{2E_{c.m.}}ln[1+exp(\dfrac{2\pi}{\hbar\omega}(E_{c.m.}-V_{B}))]
\end{equation}
If the Fusion Cross-section is described by the Wong's formula, then dimensionless quantity $F(x)$ reduces to,
\begin{equation}
F_{0}(x)=ln[1+exp(2\pi x]
\end{equation}
The quantity $F_{0}$ is known as Universal Fusion Function($UFF(x)$). 
When x$>$1, $F_{0}$(x) $\approx$ 2$\pi$x, therefore the fusion cross-section is approximated as 
\begin{equation}
\sigma_{F}=\dfrac{\pi R_{B}^{2}}{E_{c.m.}}(E_{c.m.}-V_{B})
\end{equation}
Artificial Neural Network method has been applied to extract the reliable Fusion Function $F(x)$ from the already available fusion cross-section data involving $^{6,7}$Li nuclei. Once the $F(x)$ is available, the complete fusion suppression factor due to the breakup has been estimated from the following relation,
\begin{equation}
F^{Bu}(x)=\dfrac{F(x)}{F_{0}(x)}
\end{equation}
\section*{Results and Discussions}
After conducting numerous trials in the initial phase of the study, the optimal number of hidden layer neurons was identified and employed to yield the most favorable outcomes. The empirical data for total fusion cross-sections for $^{6}$Li and $^{7}$Li nuclei, sourced from different literature~\cite{tripathi02,liu05,shrivastava09, dasgupta04,rath09,tripathi05,kumawat12,rath12,pradhan11,palshetkar14,rath13,mukherjee06,shrivastava13}, encompassed compound nuclei mass numbers ranging from 90 to 209 are used in the present analysis. The dataset comprises of dimensionless quantity $x$(independent variable) and Fusion function $F(x)$(dependent variable). Subsequently, this dataset was randomly partitioned into distinct sets for training (70$\%$) and testing (30$\%$) purposes. The neural network is implemented using scikit-learn's MLPRegressor, which stands for Multi-Layer Perceptron Regressor. After making suitable adjustments, the final weights are determined once a satisfactory level of error convergence between the predicted and desired outputs is achieved.

Upon completion of the network training, the neural network utilized the final weights to generate outputs for both training and test data values. The Normalized Mean Squared Error(NMSE) is estimated as 1.85 $\%$ for $^{6}$Li induced induced reactions and the same is 3.73 $\%$ for $^{7}$Li induced induced reactions for the training datasets. The coefficient of determination has been obtained using R$^{2}$ value, it is found to be $\approx$ 0.98 for $^{6}$Li induced reactions training data and $\approx$ 0.97 for $^{7}$Li induced reactions training data. Fig.~\ref{fig:6Li} and Fig.~\ref{fig:7Li} vividly illustrates the effective reproduction of the highly non-linear experimental fusion function $F(x)$ with dimensionless quantity $x$ by the ANN method.  

The test stage holds significant importance in confirming the success of the method. By employing the final weights of the ANN, the complete fusion function $F(x)$ in the test data for both $^{6}$Li and $^{7}$Li induced reactions were predicted. The Normalized Mean Squared Error (NMSE) for test data has been computed as 1.92$\%$ for reactions induced by $^{6}$Li and 6.48$\%$ for those induced by $^{7}$Li. R$^{2}$ value is obtained as  $\approx$ 0.98 and $\approx$ 0.95 for test data of $^{6}$Li and $^{7}$Li induced reactions respectively. 

In the final phase of the study, feed forward neural network model is used to estimate the complete fusion function value $F(x)$ for finer $x$ for both $^{6}$Li and $^{7}$Li induced reactions cases. Once the $F(x)$ is obtained, the fusion suppression factor is calculated by taking the ratio with Universal Fusion Function $F_{0}(x)$. The suppression factor is calculated as 0.68 and 0.74 for $^{6}$Li and $^{7}$Li induced reactions respectively which is quite impressive suggesting the ANN method as an alternative tool for extraction of complete fusion suppression factor. Present results has been compared with other three alternative methods Support Vector Regression, Random Forest Regression and Gaussian Process Regression. The performance matrix has been compared in the Table II and Table III. It has been observed that, ANN algorithm is giving the better accuracy as compared with other. Variation of F(x) with respect to $x$ has been manifested in Fig.~\ref{fig:svr}, Fig.~\ref{fig:rfr} and Fig.~\ref{fig:gpr} for both $^{6}$Li and $^{7}$Li induced reactions using  Support Vector Regression, Random Forest Regression and Gaussian Process Regression respectively. Subsequently, the average suppression factor has been estimated which has been tabulated in Table II and Table III. It has been observed that the ANN algorithm provides better accuracy compared to the other methods.

\section*{Conclusions}
In this study, the utilization of Artificial Neural Network (ANN) methods to estimate the complete fusion suppression factor for reactions involving $^{6,7}$Li projectiles is demonstrated. This is achieved by comparing the Universal Fusion Function ($F_{0}(x)$) with ANN-predicted reduced fusion functions ($F(x)$). The average suppression factors are found to be 0.68 and 0.74, respectively. The Normalized Mean Squared Error (NMSE) is calculated as 1.85$\%$ for training and 1.92$\%$ for testing data of ANN in the case of $^{6}$Li. For $^{7}$Li , the corresponding values are determined to be 3.73$\%$ and 6.48$\%$. The correlation coefficient, determined through the R$^{2}$ value, reveals a high degree of correlation in the study. Specifically, the correlation coefficient is approximately 0.98 for training data in $^{6}$Li induced reactions and approximately 0.97 for testing data in the same reactions. The R$^{2}$ values obtained for the test data also indicate a strong correlation, with approximately 0.98 for $^{6}$Li-induced reactions and approximately 0.95 for $^{7}$Li-induced reactions. The current results have been compared with three alternative methods: Support Vector Regression, Random Forest Regression, and Gaussian Process Regression. Relevant performance metrics have been assessed. It was observed that the ANN algorithm provides better accuracy compared to the other methods. These findings suggest that ANN might be a viable tool for estimating the complete fusion suppression factor for weakly bound projectiles.


\begin{thebibliography}{99}
\bibitem{tripathi02} V. Tripathi {\it et. al}, Phys. Rev. Lett. {\bf 88}, 172701 (2002).
\bibitem{liu05} Z. H. Liu {\it et. al}, Euro. Phys. J. A. {\bf 26}, 73 (2005).
\bibitem{shrivastava09} A. Shrivastava {\it et. al}, Phys. Rev. Lett. {\bf 103}, 232702 (2009).
\bibitem{dasgupta04} M. Dasgupta {\it et. al}, Phys. Rev. C {\bf 70}, 024606 (2004).
\bibitem{rath09} P. K. Rath {\it et. al}, Phys. Rev. C {\bf 79}, 051601(R) (2009).
\bibitem{santra11} S. Santra {\it et. al}, Phys. Rev. C {\bf 83}, 034616 (2011).
\bibitem{chattopadhyay16} D. Chattopadhyay {\it et. al}, Phys. Rev. C {\bf 94}, 061602(R) (2016).
\bibitem{chattopadhyay18} D. Chattopadhyay {\it et. al}, Phys. Rev. C {\bf 97}, 051601(R) (2018).
\bibitem{hagino00} K. Hagino {\it et. al}, Phys. Rev. C {\bf 61}, 037602 (2000).
\bibitem{diaz02} A. Diaz-Torres and I. J. Thompson, Phys. Rev. C {\bf 65}, 024606 (2002).
\bibitem{gomes11} P. R. S. Gomes {\it et. al}, Phys. Rev. C {\bf 84}, 014615 (2011).
\bibitem{kumawat12} H. KUmawat {\it et. al}, Phys. Rev. C {\bf 86}, 024607 (2012).
\bibitem{rath12} P. K. Rath {\it et. al}, Nucl. Phys. A {\bf 874}, 14 (2012).
\bibitem{pradhan11} M. K. Pradhan {\it et. al}, Phys. Rev. C {\bf 83}, 064606 (2011).
\bibitem{palshetkar14} C. S. Palshetkar {\it et. al}, Phys. Rev. C {\bf 89}, 024607 (2014).
\bibitem{rath13} P. K. Rath {\it et. al}, Phys. Rev. C {\bf 88}, 044617 (2013).
\bibitem{tripathi05} V. Tripathi {\it et. al}, Phys. Rev. C {\bf 72}, 017601 (2005).
\bibitem{canto09} L. F. Canto {\it et. al}, J. Phys. G: Nucl. Part. Phys. {\bf 36}, 015109 (2009).
\bibitem{kundu16} A. Kundu {\it et. al}, Phys. Rev. C {\bf 94}, 014603 (2016).
\bibitem{akkoyun13} S. Akkoyun {\it et. al}, Measurement {\bf 46}, 3192 (2013).
\bibitem{athanassopoulos04} S. Athanassopoulos {\it et. al}, arXiv:nucl-th/0509075, 2004.
\bibitem{athanassopoulos004} S. Athanassopoulos {\it et. al}, Nucl. Phys. A {\bf 743}, 222 (2004).
\bibitem{bayram14} T. Bayram {\it et. al}, Ann. Nucl. Energy {\bf 63}, 172 (2014).
\bibitem{david95} C. David {\it et. al}, Phys. Rev. C {\bf 51}, 1453 (1995).
\bibitem{bass96} S. A. Bass {\it et. al}, Phys. Rev. C {\bf 53}, 2358 (1996).
\bibitem{haddad97} F. Haddad {\it et. al}, Phys. Rev. C {\bf 55}, 1371 (1997).
\bibitem{akkoyun14} S. Akkoyun {\it et. al}, Pad. Phys. Chem. {\bf 96}, 186 (2014).
\bibitem{akkoyun013} S. Akkoyun {\it et. al}, J. Phys. G Nucl. Part. {\bf 40}, 055106 (2013).
\bibitem{haykin99} S. Haykin, Neural Networks: a Comprehensive Foundation, Prentice-Hall Inc., Englewood
Cliffs, NJ, USA, 1999.
\bibitem{mukherjee06} A. Mukherjee {\it et. al}, Phys. Lett. B {\bf 636}, 91 (2006).
\bibitem{shrivastava13} A. Shrivastava {\it et. al}, Phys. Lett. B {\bf 718}, 931 (2013).
\bibitem{kingma14} D. P. Kingma and J. Ba, 2014, arXiv:1412.6980.
\bibitem{nrv} http://nrv.jinr.ru/nrv/webnrv/fusion/.
\end{thebibliography}
\end{document}